\documentclass[pra,aps,groupedaddress,twocolumn,floatfix,nofootinbib]{revtex4}

\usepackage{graphicx}
\newcommand{\beq}{\begin{equation}}
\newcommand{\eeq}{\end{equation}}
\newcommand{\beqa}{\begin{eqnarray}}
\newcommand{\eeqa}{\end{eqnarray}}

\newcommand{\braket}[2]{\mbox{$ \langle #1 | #2 \rangle $}}

\newcommand{\ket}[1]{\mbox{$ | #1 \rangle $}}
\newcommand{\bra}[1]{\mbox{$ \langle #1 | $}}

\def\half{\frac{1}{2}}
\def\opone{\leavevmode\hbox{\small1\normalsize\kern-.33em1}}

\def\ptr{p_{tr}}

\begin{document}

\title{Entanglement 25 years after Quantum Teleportation:\\
testing joint measurements in quantum networks}

\author{Nicolas Gisin \\
\it \small   Group of Applied Physics, University of Geneva, 1211 Geneva 4,    Switzerland}

\date{\small \today}

\begin{abstract}
Twenty-five years after the invention of quantum teleportation, the concept of entanglement gained enormous popularity. This is especially nice to those who remember that entanglement was not even taught at universities until the 1990's. Today, entanglement is often presented as a resource, the resource of quantum information science and technology. However, entanglement is exploited twice in quantum teleportation. First, entanglement is the ``quantum teleportation channel'', i.e. entanglement between distant systems. Second, entanglement appears in the eigenvectors of the joint measurement that Alice, the sender, has to perform jointly on the quantum state to be teleported and her half of the ``quantum teleportation channel'', i.e. entanglement enabling entirely new kinds of quantum measurements. I emphasize how poorely this second kind of entanglement is understood. In particular, I use quantum networks in which each party connected to several nodes performs a joint measurement to illustrate that the quantumness of such joint measurements remains elusive, escaping today's available tools to detect and quantify it.
\end{abstract}

\maketitle

\section{Introduction}
In 1993 six co-authors surprised the world by proposing a method to teleport a quantum state from one location to a distant one \cite{Qteleportation93,QTPirandola15}. Twenty five years later the surprise is gone, but the fascination remains: How can an object submitted to the no-cloning theorem disappear here and reappear there without anything carrying any information about it transmitted from the sender, Alice, to the receiver, Bob? Today, the answer seems well known and has a name: entanglement \cite{RMPentanglement}. This merely shifts the mystery, and thus the fascination, to entanglement. However, entanglement appears twice in quantum teleportation. The first and most obvious appearance of entanglement is as the "quantum teleportation channel", i.e. entanglement between two systems, the first one controlled by Alice, the second one controlled by Bob. This sort of entanglement is by now pretty well (though no fully) understood. But entanglement appears a second time in quantum teleportation: the measurement that Alice has to perform jointly on the quantum state to be teleported and her half of the "quantum teleportation channel" has all its eigenstates maximally entangled. 

Without this second appearance of entanglement, quantum teleportation would be impossible. This can be understood intuitively as follows \cite{Qchance}. First, observe that two (maximally) entangled systems are characterized by the property that if one asks both of them the same question - i.e. perform the same measurement on each of them, then both systems deliver the same answer\footnote{As said, this is only an intuitive explanation, as there are no 2-qubit states with this property. For a more formal description of quantum teleportation, see \cite{Qteleportation93,QTPirandola15}, though the here presented intuition contains the essential points for the present context.}. Well, for singlets it's just the opposite, they get opposite results instead of identical ones, but that's just a matter of systematically flipping one of the answers. Now, the joint measurement essentially asks to two independent systems the following ``strange question'': "If I would perform the same measurement on both of you, would you provide the same answer?". This is a question about the relation between the two systems, not a pair of questions to each system whose answers are then combined in some clever way. Indeed, classical systems, including humans, can't answer such unusual joint questions. But quantum systems can. For example, the two systems can answer "yes" and get (maximally) entangled in such a way that whatever identical questions are later asked to them, they'll provide the same answer. Or the answer could be "no" and the two systems get into a different (maximally) entangled state such that their answer to arbitrary but identical questions would always be opposite. As is well-known, in order to terminate the quantum teleportation process, Alice has to communicate (classically) which result she obtained to her ``strange question''; then Bob knows whether his system will provide the same answer as had the question been asked to the original system, the one to be teleported, or whether he will receive just the opposite answer. It is important to notice that this classical communication from Alice to Bob carries exactly zero information about the teleported quantum state.

Well, in quantum theory the situation is a bit more complicated, with 4 possible answers to the joint ``strange'' measurement and a bit more involved relations between the answer and Bob's system. But the essential is there and it calls for understanding! Physics requires an understanding of such {\it joint measurements} of similar quality as our understanding of entanglement between distant systems, i.e. of entanglement as {\it quantum teleportation channels}. The quality of today's understanding of entanglement between distant systems is illustrated by its relation to Bell non-locality (i.e. Bell inequality violation) \cite{RMP-NL-Brunner14}, to quantum steering \cite{Qsteering} and, highly illuminating in my opinion, by the conceptual tool of the non-local PR-boxes that summarizes in a beautifully simple equation, $a\oplus b=x\cdot y$, the involved mathematical concept of entanglement \cite{PR94}. Something analogous for joint measurements is still missing.

\section{Quantum Teleportation and high-impact journals}
On request of the editor, let me stress that "this section presents the author's own opinion regarding publication trends in quantum information"\footnote{Let me add that this is true of all opinions expressed in all my papers.}.

Since the advent of quantum teleportation, especially since its first experimental demonstrations \cite{QTRome98,QTVienna97,QTPolzik98}, it has become quasi-mandatory to publish in journals with high impact factors, like Nature, Nature Physics, Nature Photonics, Science and PRL. For example, all papers on long-distance quantum teleportation followed that trend (well, probably I am missing some, precisely those that do not follow that pattern): \cite{LgDistQTGeneva03, QTUrsin, QTChina, QTCanaries12,QTSpace17,QTCrystal14}. So, what happens if you resist the trend? We tried. We published an experiment in which the state to be teleported was carried by a photon produced long after the entangled photons constituting the quantum teleportation channel had left the laboratory. This required that the entangled photons and the photon carrying the state to be teleported were produced by different laser pulses (though from the same laser). This appeared in J. Opt. Soc. Am. B \cite{QTSwisscom07} and received a relatively low number of citations. This is the price to pay for independence. But who cares about independence today\footnote{Here is an instructive example. I wrote (among other) the introduction to our long-distance quantum teleportation paper \cite{LgDistQTGeneva03} and cited Aristotle for his distinction of form and substance that make up objects. When the proofs arrived we discovered that the editor dared to remove all this stuff about Aristotle, form and substance (although she/he is not a co-author of our paper, isn't it?). I got angry and suggested to my students to withdraw our (accepted) submission to Nature. That proposal triggered a sort of nuclear bomb. No way to argue against the dominant fashion. I surrendered. But the arXiv version of our paper still contains Aristotle (quant-ph/0301178).}?

I am not complaining, but find it interesting to be aware of the huge impact quantum teleportation had on our community's trend to overvalue high-profile journals, with all the frustration that too often comes along. Unfortunately, that trend spread all over quantum information science. Admittedly, I am not the least responsible person for that\footnote{Though, before having students I used to send all my papers to Physics Letters A, a journal with the enormous quality of always accepting all my submissions, hence allowing me to concentrate on research.}. Sorry.

\section{The Bell-State Measurement in quantum networks}
The joint measurement exploited in quantum teleportation, known as a Bell State Measurement (BSM), is characterized by all its eigenvectors being maximally entangled. For instance, teleportation of qubits require the BSM whose eigenvectors are the four Bell states:
\beqa
\ket{\phi^\pm}&=&(\ket{0,0}\pm\ket{1,1})/\sqrt{2} \\
\ket{\psi^\pm}&=&(\ket{0,1}\pm\ket{1,0})/\sqrt{2} 
\eeqa

As already pointed out in the original paper \cite{Qteleportation93}, quantum teleportation can be extended to teleportation of entanglement, known as entanglement swapping. This, in turn, can be extended to teleportation over entire and complex networks \cite{briegel}, as illustrated in Fig.~1. In such networks, each node with more than one edge performs a joint measurement, possibly on more than two systems. For simplicity, here we concentrate on only two cases, either a line or a triangle, see Figs.~2 and 3. Notice that here only players with a single edge get inputs, denoted $x$ and $y$, that determine which measurement to perform.

Let us first consider the triangle, see Fig.~3. If Alice, Bob and Charlie each perform the BSM, then there is a simple classical model that reproduces the statistics of their outcomes, $p(a,b,c)$ - notice that there are no inputs\footnote{Another 3-partite scenario in a triangle configuration without inputs should be mentioned here \cite{TFritz12}, though it is essentially the usual CHSH 2-party case with the two random number generators collected as the third party. As expected, the resistance to noise per singlet is poor, certainly not better than for the usual CHSH inequality.}. Hence, somewhat surprisingly, in this case the joint measurement doesn't produce any quantum signature: such a triangle with BSM displays no quantumness.

Let's now consider the line of Fig.~2. Start with only 2 edges. This corresponds to the scenario of entanglement swapping, i.e. of quantum teleportation of entanglement. For this simple case we name the parties with their usual names, i.e. Alice, Bob and Charlie, instead of $A_1$, $A_2$ and $A_3$, and similarly for the outcomes. Depending on Bob's outcome $b$, Alice's and Charlie's qubits get projected onto different entangled states; which exact entangled state depends on $b$. This can be checked with some entanglement witness, or, in a device-independent way, with some Bell inequality. For the CHSH inequality, assuming perfect (noise-free) measurements, a violation is obtained if the product of the visibilities\footnote{Recall that for the Werner states $\rho_W=W\cdot\ket{\psi^-}\bra{\psi^-}+(1-W)\opone/4$, where $\ket{\psi^-}$ denotes the singlet, the visibility equals $W$.} satisfies $W_1\cdot W_2>1/\sqrt{2}$. In the symmetric case, $W_1=W_2$, which implies $W_j>2^{-1/4}\approx 84\%$. Such a high visibility has been achieved experimentally, e.g. \cite{EntSwapNL02}, but with non-independent sources for the two quantum states $\rho_1$ and $\rho_2$ represented by the edges.
\begin{figure} 
\centering
\includegraphics[width=10cm]{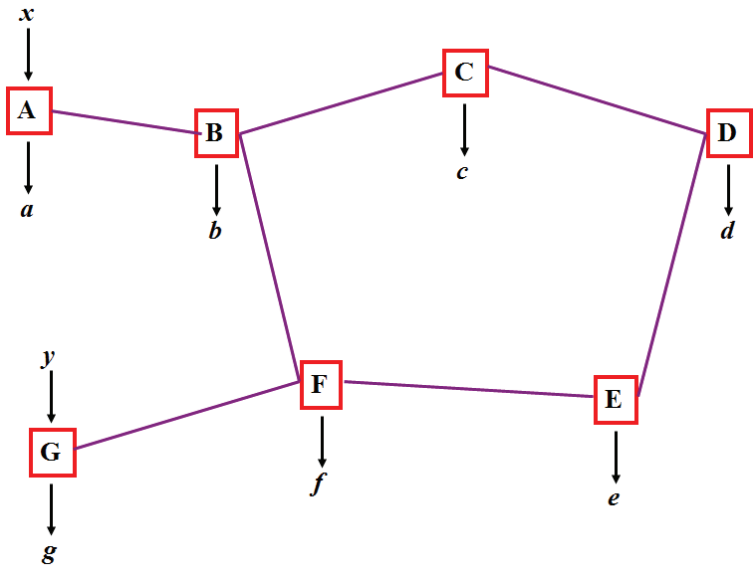}
\vspace{-10pt}
\caption{\it Example of a quantum network. Each edge represents a resource shared by the connected nodes. The resource are entangled quantum states, or, in order to compare with classical networks, correlated local variables (i.e. shared randomness). In this paper we consider only cases where inputs are provided to parties connected by a single edge.}
\end{figure}

\begin{figure}
\centering
\includegraphics[width=6cm]{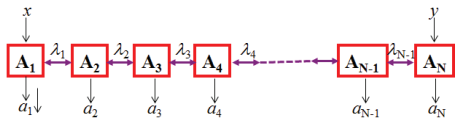}
\vspace{-10pt}
\caption{\it (N-1)-local scenario in a line \cite{triangle}. The $\lambda_j$'s represent independent quantum states, or, in the classical scenario used for comparison, random independent local variables. Only the first and last parties get inputs, $x$ and $y$ respectively.}
\end{figure}

\begin{figure} 
\centering
\includegraphics[width=6cm]{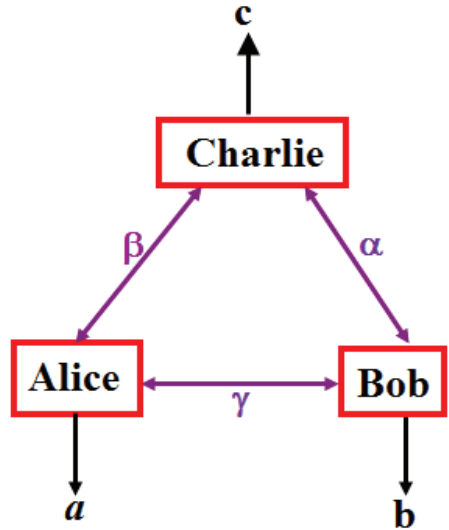}
\vspace{-10pt}
\caption{\it The triangle configuration for 3 parties \cite{triangle}. Each pair of parties shares either a quantum state and performs quantum measurements - quantum scenario, or shares independent random variables $\alpha$, $\beta$ and $\gamma$ and outputs a function of the random variables to which they have access. Notice that the three random variables are only used locally, hence the terminology 3-local scenario. The ``Quantum Grail'' is to find a quantum scenario (without external inputs) leading to a probability $p(a,b,c)$ which can't be reproduced by any 3-local scenario.}
\end{figure}

However, in such an entanglement scenario with independent sources, like e.g. \cite{Matthaeus07}, it is very natural to check for quantumness by comparing it with classical correlations under the assumption that the local (hidden) variables are also independent:
\beq
P(\lambda_1, \lambda_2)=P(\lambda_1)\cdot P(\lambda_2)
\eeq
Such a case is called bi-local \cite{bilocality10,bilocality12}, to contrast it with the usual Bell locality. In case of $n$ independent sources, the achievable classical correlations are called $n$-local \cite{Nloc16, Bell2Nloc, Armin16}.

In the bi-locality scenario it has been proven that a visibility product of $W_1\cdot W_2>\half$ suffices to prove quantumness, i.e. to prove a quantum advantage over bi-local classical correlations \cite{bilocality10,bilocality12}. Accordingly, in the symmetric case $W_j<1/\sqrt{2}\approx 71\%$ suffices, as, e.g., in the experiment of Ref.~\cite{Matthaeus07}. In this scenario, an explicit non-linear inequality (non-linear because the set of $n$-local correlations in non convex for all $n\geq2$) has been found and fully analysed \cite{bilocCHSH17}. The analyses show that this bi-local scenario is essentially identical to the old and well-known CHSH Bell inequality between 2 parties. The relation builds on the fact that the 2-bit outcome of the BSM is equivalent to the outcome of $\sigma_z\otimes\sigma_z$ for the first bit and $\sigma_x\otimes\sigma_x$ for the second bit. Hence, in a nutshell, Bob measures both of his qubits in the $x-z$ bases, while Alice and Charlie measure in the $\pm45^\circ$ bases, exactly as in the CHSH case.

This is quite disappointing, as the threshold visibility per singlet, $1/\sqrt{2}$, is identical to the simpler case of CHSH between 2 parties. Apparently, the assumption of independent local variables $\lambda_1$ and $\lambda_2$ plays no role. But that cannot be! Independence is a strong assumption, it should thus lead to consequences. This illustrates how poorly we understand joint measurements. Could it be that increasing the number of inputs at Alice and Charlie's side, or studying longer linear chains, allows one to lower the threshold visibility per singlet? Reference \cite{Nloc16}, which considers $n$-locality in longer lines, and reference \cite{Bell2Nloc}, which derives $n$-local inequalities from Bell inequalities, suggest the contrary and, so far, numerous numerical searches lead to disappointing results, see though the interesting findings in \cite{Chaves16, Wolfe18, Fritz16, Sciarrino17}.

The mentioned negative results are no proof that the bi-local scenario is useless to lower the threshold visibility per singlet. But they call for alternative ideas. One nice idea is to go for a star network \cite{tavakoli14,Bell2Nloc}, though so far results seem very similar to the bi-local case. 

The next section recalls results first presented in \cite{TriangleQuantph}, a paper I never submitted to any journal, hence parts of it are reproduced here. In a nutshell, it presents another joint measurement and applies it to a 3-partite scenario in the triangle configuration with 3 independent sources.

\section{The Elegant Joint Measurement on 2 qubits}
In order to study joint measurements different from the BSM we like to find a 2-qubit basis with 4 partially entangled eigenstates, all with the same degree of entanglement and some nice symmetries. For this, we start with the 4 vertices of the tetrahedron inscribed in the poincar\'e sphere:
\beqa
\vec m_1&=& (1,1,1)/\sqrt{3} \label{m1} \\
\vec m_2&=& (1,-1,-1)/\sqrt{3} \label{m2} \\
\vec m_3&=& (-1,1,-1)/\sqrt{3} \label{m3} \\
\vec m_4&=& (-1,-1,1)/\sqrt{3} \label{m4} 
\eeqa

Using cylindrical coordinates, $\vec m_j=(\sqrt{1-\eta_j^2}\cos{\phi_j}, \sqrt{1-\eta_j^2}\sin{\phi_j},\eta_j)$, one obtains the natural correspondence with qubit states (note that here $\eta_j=\pm1/\sqrt{3}$ for all $j$):
\beq\label{ketm}
\ket{\vec m_j}=\sqrt{\frac{1-\eta_j}{2}}e^{i\phi_j/2}\ket{0}+\sqrt{\frac{1+\eta_j}{2}}e^{-i\phi_j/2} \ket{1}
\eeq
Note that $\vec m_j=\bra{\vec m_j}\vec\sigma\ket{\vec m_j}$, as expected (with $\vec\sigma$ the 3 Pauli matrices). \\

Inspired by \cite{MassarPopescu,antiParallelSpins}, we consider the following 2-qubit basis constructed on anti-parallel spins  \cite{TriangleQuantph}:
\beqa\label{Psij}
\ket{\Phi_j}&=&\sqrt{\frac{3}{2}}\ket{\vec m_j, -\vec m_j} + i\frac{\sqrt{3}-1}{2}\ket{\psi^-} \\
&=&\frac{\sqrt{3}+1}{2\sqrt{2}}\ket{\vec m_j,-\vec m_j}+\frac{\sqrt{3}-1}{2\sqrt{2}}\ket{-\vec m_j,\vec m_j} \label{SchmidtBasis}
\eeqa
where $\ket{-\vec m}$ is orthogonal to $\ket{\vec m}$: it has the same form as (\ref{ketm}) but with $\eta\rightarrow -\eta$ and $\phi\rightarrow\phi+\pi$. Notice that in (\ref{SchmidtBasis}) the states $\Phi_j$ are written in their Schmidt bases.

In order to check that the $\Phi_j$ are normalised and mutually orthogonal one should use $\braket{\vec m, -\vec m}{\psi^-}=i/\sqrt{2}$ for all $\vec m$ and $\braket{\vec m_j, -\vec m_j}{\vec m_k, -\vec m_k}=1/3$ for all $j\neq k$.

Using the corresponding 1-dimensional projectors:
\beqa
\ket{\Phi_j}\bra{\Phi_j}&=&\frac{1}{4}\left(\opone+\frac{\sqrt{3}}{2}(\vec m_j\vec\sigma\otimes\opone - \opone\otimes\vec m_j\vec\sigma) 
-\frac{3}{2}\sum_{n,k}m_{j,n}m_{j,k}\sigma_n\otimes\sigma_k + \half\vec\sigma\otimes\vec\sigma \right)
\eeqa
it is not difficult to compute the partial traces and observe the elegant properties:
\beqa
\bra{\Phi_j}\vec\sigma\otimes\opone\ket{\Phi_j}&=&\half\vec m_j \\
\bra{\Phi_j}\opone\otimes\vec\sigma\ket{\Phi_j}&=&-\half\vec m_j
\eeqa
In words, the partial states (obtained by tracing out one party) point along the edges of the tetrahedron, but with Bloch vectors of reduced lengths $\half$. \\

We name the 2-qubit measurement with eigenstates (\ref{Psij}-\ref{SchmidtBasis}) the Elegant Joint Measurement (EJM). We believe it is unique with all 4 eigenstates having identical degrees of partial entanglement and with all partial states of all eigenstates parallel or anti-parallel to the vertices of the tetrahedron.

\section{Quantum Correlation from singlets and the EJM in the triangle configuration}
Consider 3 independent singlets in the triangle configuration and assume that Alice, Bob and Charlie each perform the EJM on their 2 (independent) qubits, see Fig.~3. Denote the resulting correlation $\ptr(a,b,c)$, where $a,b,c=1,2,3,4$. By symmetry, $\ptr(a,b,c)$ is fully characterized by 3 numbers corresponding to the cases $a=b=c$, $a=b\ne c$ and $a\ne b\ne c\ne a$ (and circular permutations, i.e. 2 outcomes are equal, but the third differs). A not too complex computation gives \cite{TriangleQuantph}:
\beqa
\ptr(a=k,b=k,c=k)&=&\frac{25}{256}~ for~ k=1,2,3,4 \label{correl1}\\
\ptr(a=k,b=k,c=m)&=&\frac{1}{256}~ for~ k\ne m  \label{correl2}\\
\ptr(a=k,b=n,c=m)&=&\frac{5}{256}~ for~ k\ne n\ne m\ne k  \nonumber\\ \label{correl3}
\eeqa
The normalization holds: $4\cdot\frac{25}{256}+36\cdot\frac{1}{256}+24\cdot\frac{5}{256}=1$.

As expected $\ptr(a)=\ptr(b)=\ptr(c)=\frac{1}{4}$. More interesting is the probabilities that two parties get identical results:
\beqa
\ptr(a=k,b=k)&=&\ptr(a=b=c=k)+\ptr(a=b=k,c\ne k)\nonumber\\
&=&\frac{25+3\cdot 1}{256}=\frac{7}{64}
\eeqa
Hence, all pairs of parties are correlated, e.g. $\ptr(a|b)\ne\frac{1}{4}$. In worlds, given an outcome $b=k$ for Bob, Alice's outcome has a large chance to take the same value: $\ptr(a=k|b=k)=\frac{\ptr(a=k,b=k)}{\ptr(b=k)}=\frac{7}{16}$. Accordingly:
\beq
\ptr(a=b)=\sum_k \ptr(b=k)p(a=k|b=k)=\frac{7}{16}
\eeq

The strength of the 3-party correlation is even more impressive:
\beq
\ptr(a=k|b=c=k)=\frac{\ptr(a=b=c=k)}{\ptr(b=c=k)}=\frac{25}{28}
\eeq
Hence $\ptr(a=b=c)=4\cdot\frac{25}{256}=\frac{25}{64}$.

The high correlation displayed by $\ptr$ strongly suggests that it can't be realized by any 3-local model. However, one has to be careful. Indeed, reference \cite{TriangleQuantph} presents two 3-local models with even higher correlations, though not symmetric and not reproducing the correlations (\ref{correl1}-\ref{correl3}) of $\ptr$. For completeness, these two models are reproduced in the next section \ref{ptr}. Since \cite{TriangleQuantph} was posted on the arXiv quite some researchers tried to prove or disprove the 3-local nature of $\ptr$. In particular Elisa B\"aumer and Elie Wolfe (private communications) devoted time to this fascinating question, the first one with strong arguments in favour of a negative answer and the second one, using his ``inflation method'' \cite{WolfeInflation,NavascuesWolfeInflation}, arguing in favour of a positive answer. The fact is that the 3-local nature of $\ptr$ remains elusive. More generally, the existence/nonexistence of a quantum scenario that can provably not be reproduced by any 3-local model and that respects the triangle symmetry, or some other closed symmetric loop, remains open, illustrating how poorely we understand joint measurements. Let me emphasize that if such a quantum example exists, its quantumness could only be due to the joint measurements, as in a loop there are no ``ends'', hence no parties with inputs, in strong contrast to the by now common Bell inequality scenarios. I elaborate on this in section \ref{consQtriangle}.

\section{Is {\boldmath $\ptr(a,b,c)$} 3-local?}\label{ptr}
In this section, we consider the question whether the quantum probability $\ptr(a,b,c)$ is 3-local, i.e. whether it can be reproduced by a 3-local model:
\beq\label{3loc}
\ptr\stackrel{?}{=}  \sum_{\alpha\beta\gamma} P(\alpha)P(\beta)P(\gamma) P(a|\beta,\gamma)P(b|\gamma,\alpha)P(c|\alpha,\beta)
\eeq

In such a 3-local model of $\ptr(a,b,c)$ the Alice-Bob correlation could only be due to their shared randomness $\gamma$. Similarly, the correlation between Bob and Charlie is necessarily due to $\alpha$ and the Alice-Charlie correlation due to $\beta$. Accordingly, each local variable $\alpha$, $\beta$ and $\gamma$ would contain a 4-dit, equally distributed among the values 1,2,3,4, and with a relatively high probability both Alice and Bob output the 4-dit contained in $\gamma$, and similarly for the other pairs of parties. Admittedly, this is only an argument, not a proof of the conjecture that $\ptr$ is non-local.

Accordingly, let's consider the following natural type of 3-local models. Let $\gamma=(\gamma_1,\gamma_2)$, where $\gamma_1=1,2,3,4$, each with equal probability and $\gamma_2=0,1$ with $prob(\gamma_2=1)=q$. The idea is that whenever $\gamma_2=1$, then Alice and Bob results are given by $\gamma_1$, hence Alice and Bob get perfectly correlated. More explicitly, Alice's output function reads:
\beq\label{abetagamma}
a(\beta,\gamma)=\left\{
\begin{array}{c}
	\hspace{0.5cm}\gamma_1\hspace{5mm} if \hspace{2mm}\beta_2=0\hspace{2mm} and \hspace{2mm}\gamma_2=1\\
  \hspace{0.5cm}\beta_1\hspace{0.5cm}if~\beta_2=1\hspace{2mm}and\hspace{2mm}\gamma_2=0 \\
  \beta_1|\gamma_1\hspace{2mm}if\hspace{2mm}\beta_2=\gamma_2\hspace{15mm}
\end{array}
\right.
\eeq
where $\beta_1|\gamma_1$ indicates that $a(\beta,\gamma)$ equals $\beta_1$ or $\gamma_1$ with equal probability $\half$.

Table I indicates all possible outputs (where $\bar q\equiv(1-q)=prob(\alpha_2=0)=prob(\beta_2=0)=prob(\gamma_2=0))$.
\begin{table}[h]
	\centering
		\begin{tabular}
			{c|c|c|c|c|c|c|c|c}
			\large
			$\alpha_2$&$\beta_2$&$\gamma_2$&a&b&c&P&prob(a=b)&prob(a=b=c)\\
			\hline
			0&0&0&$\beta_1|\gamma_1$&$\alpha_1|\gamma_1$&$\alpha_1|\beta_1$&$\bar q^3$&7/16&13/64 \\
			0&0&1&$\gamma_1$&$\gamma_1$&$\alpha_1|\beta_1$&$\bar q^2q$&1&1/4 \\
			0&1&0&$\beta_1$&$\alpha_1|\gamma_1$&$\beta_1$&$\bar q^2q$&1/4&1/4 \\
			0&1&1&$\beta_1|\gamma_1$&$\gamma_1$&$\beta_1$&$\bar q q^2$&5/8&1/4 \\
			1&0&0&$\beta_1|\gamma_1$&$\alpha_1$&$\alpha_1$&$\bar q^2q$&1/4&1/4 \\
			1&0&1&$\gamma_1$&$\alpha_1|\gamma_1$&$\alpha_1$&$\bar q q^2$&5/8&1/4 \\
			1&1&0&$\beta_1$&$\alpha_1$&$\alpha_1|\beta_1$&$\bar q q^2$&1/4&1/4 \\
			1&1&1&$\beta_1|\gamma_1$&$\alpha_1|\gamma_1$&$\alpha_1|\beta_1$&$q^3$&7/16&13/64 \\
		\end{tabular}
\caption{\it The 8 lines correspond to the 8 possible combinations of values of $\alpha_2$, $\beta_2$ and $\gamma_2$ (first 3 columns). The next 3 columns indicate Alice, Bob and Charlie's outputs. The 7th column indicates the probability of the corresponding line and the last two columns the probability that $a=b$ and $a=b=c$, respectively.}
\end{table}

Averaging the probabilities that $a=b=c$ over the 8 combinations of values of $\alpha_2$, $\beta_2$ and $\gamma_2$, i.e. over the 8 lines of Table 1, gives:
\beqa
p_{3loc}(a=b=c)&=&\frac{13}{64}(\bar q^3+q^3)+\frac{3}{4}(\bar q^2q+\bar q q^2) \nonumber\\
&=&\frac{13+9q-9q^2}{64}
\eeqa
Hence, the maximal 3-partite correlation of our 3-local model is achieved for $q=\half$ and reads:
\beq
\max_q~p_{3loc}(a=b=c)=\frac{61}{256}
\eeq
This is much smaller than the value obtained in the quantum case with the Elegant Joint Measurement.

The above is not a proof, but leads us to conjecture that the quantum probability $\ptr(a,b,c)$ is not 3-local. Indeed, $\gamma$ has to correlate A an B, i.e. $\gamma$ contributes to the probability that $a=b$, and $\beta$ contributes to $\ptr(a=c)$ and $\alpha$ contributes to $\ptr(b=c)$. But then the three independent variables $\alpha$, $\beta$ and $\gamma$ can't do the job for the 3-partite correlation $a=b=c$.

Note that if the outcomes are grouped 2 by 2, such that outcomes are binary, then a 3-local model similar to (\ref{abetagamma}) can reproduce the quantum correlation. But, again, with 4 outcomes per party this seems impossible.

\subsection{A natural but asymmetric 3-local model}
There is another 3-local model that we need to consider, directly inspired by the quantum singlet states shared by each pair of parties. Assume that the three local variables $\alpha$, $\beta$ and $\gamma$ each take values (0,1) or (1,0) with 50\% probabilities, where the first bit of $\alpha$ is sent to Bob and the second bit to Charlie, and similarly for $\beta$ and $\gamma$. Clearly, this 3-local model assumes binary local variables, i.e. bits,  but we like to keep the notation (0,1) and (1,0) for the two values.

The outcomes are then determined by the two bits that each party receives from the local variables it shares with his two neighbours. We like to maximize the probability $p(a=b=c)$. All output functions that maximize $p(a=b=c)$ are equivalent. One possible choice  is:
\beqa
(0,0)\Rightarrow a=2,~b=4,~c=3~~ \\
(0,1)\Rightarrow a=1,~b=1,~c=1~~ \\
(1,0)\Rightarrow a=3,~b=2,~c=4~~ \\
(1,1)\Rightarrow a=4,~b=3,~c=2~~
\eeqa
Note that in this 3-local model $\gamma$ imposes that both Alice and Bob can only output one out of two values. Which of the two values happens depends on the second local variable. This provides intuition why this 3-local model achieves $p(a=b=c)=\half$, i.e. an even larger value than the quantum probabilities with the EJM. Moreover $p(a=b)=\half$, hence $p(a=b=c|a=b)=1$. However, this model does not respect the symmetries of the quantum scenario. In particular 20 out of the 24 cases $p(a=k,b=n,c=m)$ with $k\ne n\ne m\ne k$ take values 0 (recall that in the quantum scenario all 24 probabilities take value $\frac{5}{256}$, see eq. (\ref{correl1}-\ref{correl3})).

This simple 3-local model shows that in order to prove the non-3-locality of $\ptr(a,b,c)$ it is not sufficient to consider $p(a=b=c)$, but one has to consider also the cases $a\ne b\ne c$.

\section{Consequences of a non-3-local quantum triangle}\label{consQtriangle}
Let's assume that there is a nicely symmetric quantum example of a triangle provably not 3-local, e.g. a probability distribution $p(a,b,c)$ which derives from 3 independent quantum states and identical quantum measurements in the triangle configuration, see Fig.~3, that has no 3-local decomposition (\ref{3loc})\footnote{While finishing this work, Prof. Salman Beigi sent me what appears to be the first such example \cite{Salman}!. For a non-symmetric example see Fritz exaple \cite{TFritz12} recalled in footnote 5.}. What would that imply for our worldview? First, notice that in such a scenario there are no inputs. Accordingly, one could imagine a toy universe consisting of only 6 qubits, without anything outside, which nevertheless manifests quantumness, including provable randomness. Well, the outcomes $a$, $b$ and $c$ should get out of this mini-quantum-universe in order to produce any evidence; one more manifestation of the infamous quantum measurement problem \cite{CollapseGisin18,Found2App}. This is in strong contrast to the usual Bell inequality scenario where inputs provided from outside the systems under test are essential to prove any quantumness. Of course, our 6 qubit toy universe must satisfy the assumption of independence of the 3 sources (without any assumption, nothing can be proven). But this assumption is really minimal: if the sources are spatially separated, then it is very natural to assume that they are independent. The first source could be powered by solar power and produce entangled photons, the second source powered by human energy and produce entangled atoms, and the third source powered by nuclear power and produce some entangled quantum "stuff", e.g. cats or crystals \cite{EntCrystal}.

Admittedly, one may argue that Alice, for instance, somehow gets inputs from the sources denoted $\beta$ and $\gamma$ on Fig.~3. But in Bell inequality scenarios one never thinks of the source in-between Alice and Bob as the inputs, the inputs are determining the measurement setting and, in Bell scenarios, necessarily come from outside the quantum systems. Nothing like this in the triangle scenario. Quantumness would be proven from inside the 6 qubit toy universe\footnote{Note that the 6 qubits could also be on a line, as in Fig.~7 of \cite{bilocality12}.}. Also quantum randomness would be proven within this toy universe.

A second interesting consequence of a "quantum triangle" appears when one moves the sources $\alpha$, $\beta$ and $\gamma$ close to one of the players, or even inside the players. Assume the source $\alpha$ is given to Bob, $\beta$ is given to Charlie and $\gamma$ to Alice. In the quantum case, Alice, Bob and Charlie each emits some quantum state, e.g. one qubit, and sends it to his partner counter-clock wise. In the classical case they each send an arbitrarily large amount of classical information (possibly infinite) to their partner, still counter-clock wise. The 3-local assumption of independence translates into the assumption that all communications are well enough synchronized to guarantee that each party sends out his quantum state or classical information before receiving anything from his partner. In this way one compares the power of quantum communication (of even just a qubit) with the power of classical communication, possibly an infinite amount of classical information. Under the synchronization assumption of the communications, one would prove the superiority of the former over the latter.

Admittedly, a similar story of replacing entanglement (shared randomness) by quantum (classical) communication can be told for the standard Bell inequality scenario. Instead of an entanglement source in-between Alice and Bob, Alice would send a quantum state to Bob prior to receiving her input $x$. This would allow them to violate the CHSH-Bell inequality, while if Alice is restricted to sending classical information - prior to receiving her input - they can't violate any Bell inequality.

\section{Conclusion}
In summary, 25 years after the beautiful invention of quantum teleportation lots of progress has been made on Bell-locality \cite{RMP-NL-Brunner14}, on quantum steering \cite{Qsteering} and more generally quantum information theory. Likewise enormous progress happens in experimental, applied and engineering, even in industrialization of quantum technologies \cite{QKD-IDQ,QRNG-SKT,QKD421km}. But, quite surprisingly and disappointingly, essentially no progress took place in improving our understanding of joint measurements\footnote{One exception is the possibility of detecting joint measurements in a device-independent way, see, e.g., \cite{DIJMBrunner, DIJMLee, DIJMRenou, DIJMBancal}}, i.e. on the second usage of entanglement in quantum teleportation. For example, it was proven that there is no simple analog of PR-boxes for joint measurements \cite{noCouplers,noCouplerJB, BrunnerSkrypczyk1,BrunnerSkrypczyk2}. This is exciting, as it indicates that big surprises still await us in the - hopefully not too far - future.

\small{
\section*{Acknowledgement}
The present version profited from valuable comments by Nicolas Brunner, Sandu Popescu, Armin Tavakoli and - for once - the two referees. Work partially supported by the Swiss NCCR-QSIT and the European ERC-AG MEC.
}

\end{document}